\global\def\draftcontrol{0}

   \def\versionno{Deformation of the cone  over $Y^{p,q}$}

\catcode`\@=11

\expandafter\ifx\csname draftcontrol\endcsname\relax\global\def\draftcontrol{0}
\fi

{\count255=\time\divide\count255 by 60
\xdef\hourmin{\number\count255}
\multiply\count255 by-60\advance\count255 by\time
\xdef\hourmin{\hourmin:\ifnum\count255<10 0\fi\the\count255}}
\def\draftdate{\number\month/\number\day/\number\year\ \ \ \hourmin }

\newcommand\makepapertitle{\par
  \begingroup
    \renewcommand\thefootnote{\@fnsymbol\c@footnote}%
    \def\@makefnmark{\rlap{\@textsuperscript{\normalfont\@thefnmark}}}%
    \long\def\@makefntext##1{\parindent 1em\noindent
            \hb@xt@1.8em{%
                \hss\@textsuperscript{\normalfont\@thefnmark}}##1}%
     \newpage
     \global\@topnum\z@   
     \@makepapertitle
     \thispagestyle{empty}\@thanks
  \endgroup
  \setcounter{footnote}{0}%
  \global\let\thanks\relax
  \global\let\makepapertitle\relax
  \global\let\@makepapertitle\relax
  \global\let\@thanks\@empty
  \global\let\@author\@empty
  \global\let\@date\@empty
  \global\let\@title\@empty
  \global\let\title\relax
  \global\let\author\relax
  \global\let\date\relax
  \global\let\and\relax
  \def\version{\let\version\@version\@gobble}
}
\def\@makepapertitle{%
  \newpage
   \ifnum\draftcontrol=1 {}
   \version\versionno
   \vskip 3em%
   \else
   \hfill\hbox to 3cm {\parbox{4cm}{\@pubnum}\hss}%
   \vskip 3em%
   \fi
   \begin{center}%
   \let \footnote \thanks
     {\LARGE {\@title}}%
     \vskip 1.5em%
     {\normalsize
       \lineskip .5em%
       \begin{tabular}[t]{c}%
         \@author
       \end{tabular}\par}%
     \vskip 1.5em%
     {\@bstract}%
     \end{center}%
     \vskip 1.5em
     \@date%
   \par
}

\gdef\@pubnum{}
\def\pubnum#1{%
  \gdef\@pubnum{#1}}

\gdef\@bstract{}
\def\Abstract#1{%
  \gdef\@bstract{%
   \parbox{\textwidth-0pc}{%
   \centerline{\bf Abstract}\penalty1000%
\noindent
\renewcommand\baselinestretch{1.0}%
{#1}}}
}

\def\ps@paper{\let\@mkboth\@gobbletwo%
     \ifnum\draftcontrol=1
        \def\@oddfoot{\hbox to \textwidth{\tiny \versionno \hfil\tiny\draftdate}%
        \hskip -\textwidth \hbox to \textwidth{\hfil\rm\thepage\hfil}}%
     \else\def\@oddfoot{\hbox to \textwidth{\hfil\rm\thepage\hfil}}
     \fi
     \let\@evenfoot\@oddfoot
}

\def\@version#1{\ifnum\draftcontrol=1
\typeout{}\typeout{#1}\typeout{}
\vskip3mm\centerline{\hbox{\fbox{\normalsize{\tt DRAFT -- #1 -- }
                   {\draftdate}}}}\vskip3mm
\fi}
\let\version\@version
\long\def\eqlabel#1{\ifnum\draftcontrol=1
                    \tag@false  
                    \tag*{(\theequation) \hbox to -0.2cm{\hspace{0cm}\small{#1}\hss}}
                    \refstepcounter{equation}
                    \edef\@currentlabel{\theequation}
                    \ltx@label{#1}          
                    \else
                    \label{#1}
                    \fi
                    }
\let\st@bibitem\@bibitem
\let\st@lbibitem\@lbibitem
\ifnum\draftcontrol=1
  \def\@bibitem#1{%
    \st@bibitem{#1}\a@@label{#1}\ignorespaces}
  \def\@lbibitem[#1]#2{%
    \st@lbibitem[#1]{#2}\a@@label{#2}\ignorespaces}
  \def\a@@label#1{%
    \gdef\a@lab{\smash{\normalfont\small#1}}
    \ifvmode
      \if@inlabel
        \global\setbox\@labels\hbox{%
          \llap{\a@lab\let\a@lab\relax
                \kern\@totalleftmargin\kern\marginparsep}%
          \box\@labels}%
      \fi
    \fi}
\fi

\documentclass[12pt,letterpaper]{article}

\usepackage{amsmath,amssymb,array,calc,rotating,epsfig,psfrag}
\usepackage[nosort]{cite}

\ifnum\draftcontrol=1
\fi

\renewcommand\baselinestretch{1.25}
\renewcommand\section{\@startsection {section}{1}{\z@}%
                                   {-3.5ex \@plus -1ex \@minus -.2ex}%
                                   {2.3ex \@plus.2ex}%
                                   {\normalfont\large\bfseries}}
\renewcommand\subsection{\@startsection{subsection}{2}{\z@}%
                                   {-3.25ex\@plus -1ex \@minus -.2ex}%
                                   {1.5ex \@plus .2ex}%
                                   {\normalfont\normalsize\bfseries}}
\renewcommand\subsubsection{\@startsection{subsubsection}{3}{\z@}%
                                   {-3.25ex\@plus -1ex \@minus -.2ex}%
                                   {1.5ex \@plus .2ex}%
                                   {\normalfont\normalsize\it}}
\renewcommand\paragraph{\@startsection{paragraph}{4}{\z@}%
                                   {-3.25ex\@plus -1ex \@minus -.2ex}%
                                   {1.5ex \@plus .2ex}%
                                   {\normalfont\normalsize\bf}}

\def\revise#1       {\raisebox{-0em}{\rule{3pt}{1em}}%
                     \marginpar{\raisebox{.5em}{\vrule width3pt\
                     \vrule width0pt height 0pt depth0.5em
                     \hbox to 0cm{\hspace{0cm}{%
                     \parbox[t]{4em}{\raggedright\footnotesize{#1}}}\hss}}}}

\def\calf         {{\cal F}}

\def\del          {\partial}

\def\tr           {\mathop{\rm Tr}}

\def\half{{\frac12}}

\def\sqr#1#2{{\vcenter{\vbox{\hrule height.#2pt
 \hbox{\vrule width.#2pt height#1pt \kern#1pt
 \vrule width.#2pt}\hrule height.#2pt}}}}

\newcommand{\fft}[2]{{\frac{#1}{#2}}}
\newcommand{\ft}[2]{{\textstyle{\frac{#1}{#2}}}}

\def\b{\beta}
\def\r{\rho}

\def\m{\mu}
\def\g{\gamma}

\def\n{\nu}
\def\bn{\bar{\nu}}
\def\bm{\bar{\mu}}

\catcode`\@=12

\usepackage{color}

\begin{document}




\newcommand{\be}{\begin{equation}}
\newcommand{\ee}{\end{equation}}
\newcommand{\beq}{\begin{equation}}
\newcommand{\eeq}{\end{equation}}
\newcommand{\ba}{\begin{eqnarray}}
\newcommand{\ea}{\end{eqnarray}}
\newcommand{\nn}{\nonumber}

\def\vol{\bf vol}
\def\Vol{\bf Vol}
\def\del{{\partial}}
\def\vev#1{\left\langle #1 \right\rangle}
\def\cn{{\cal N}}
\def\co{{\cal O}}
\def\IC{{\mathbb C}}
\def\IR{{\mathbb R}}
\def\IZ{{\mathbb Z}}
\def\RP{{\bf RP}}
\def\CP{{\bf CP}}
\def\Poincare{{Poincar\'e }}
\def\tr{{\rm tr}}
\def\tp{{\tilde \Phi}}
\def\Y{{\bf Y}}
\def\te{\theta}
\def\bX{\bf{X}}

\def\TL{\hfil$\displaystyle{##}$}
\def\TR{$\displaystyle{{}##}$\hfil}
\def\TC{\hfil$\displaystyle{##}$\hfil}
\def\TT{\hbox{##}}
\def\HLINE{\noalign{\vskip1\jot}\hline\noalign{\vskip1\jot}} 
\def\seqalign#1#2{\vcenter{\openup1\jot
  \halign{\strut #1\cr #2 \cr}}}
\def\lbldef#1#2{\expandafter\gdef\csname #1\endcsname {#2}}
\def\eqn#1#2{\lbldef{#1}{(\ref{#1})}%
\begin{equation} #2 \label{#1} \end{equation}}
\def\eqalign#1{\vcenter{\openup1\jot
    \halign{\strut\span\TL & \span\TR\cr #1 \cr
   }}}
\def\eno#1{(\ref{#1})}
\def\href#1#2{#2}
\def\half{{1 \over 2}}

\def\ads{{\it AdS}}
\def\adsp{{\it AdS}$_{p+2}$}
\def\cft{{\it CFT}}

\newcommand{\ber}{\begin{eqnarray}}
\newcommand{\eer}{\end{eqnarray}}

\newcommand{\bea}{\begin{eqnarray}}
\newcommand{\eea}{\end{eqnarray}}

\newcommand{\beqar}{\begin{eqnarray}}
\newcommand{\cN}{{\cal N}}
\newcommand{\cO}{{\cal O}}
\newcommand{\cA}{{\cal A}}
\newcommand{\cT}{{\cal T}}
\newcommand{\cF}{{\cal F}}
\newcommand{\cC}{{\cal C}}
\newcommand{\cR}{{\cal R}}
\newcommand{\cW}{{\cal W}}
\newcommand{\eeqar}{\end{eqnarray}}
\newcommand{\lm}{\lambda}\newcommand{\Lm}{\Lambda}
\newcommand{\eps}{\epsilon}


\newcommand{\nonu}{\nonumber}
\newcommand{\oh}{\displaystyle{\frac{1}{2}}}
\newcommand{\dsl}
  {\kern.06em\hbox{\raise.15ex\hbox{$/$}\kern-.56em\hbox{$\partial$}}}
\newcommand{\as}{\not\!\! A}
\newcommand{\ps}{\not\! p}
\newcommand{\ks}{\not\! k}
\newcommand{\D}{{\cal{D}}}
\newcommand{\dv}{d^2x}
\newcommand{\Z}{{\cal Z}}
\newcommand{\N}{{\cal N}}
\newcommand{\Dsl}{\not\!\! D}
\newcommand{\Bsl}{\not\!\! B}
\newcommand{\Psl}{\not\!\! P}
\newcommand{\eeqarr}{\end{eqnarray}}
\newcommand{\ZZ}{{\rm \kern 0.275em Z \kern -0.92em Z}\;}

\def\s{\sigma}
\def\a{\alpha}
\def\b{\beta}
\def\r{\rho}
\def\d{\delta}
\def\g{\gamma}
\def\G{\Gamma}
\def\ep{\epsilon}
\makeatletter \@addtoreset{equation}{section} \makeatother
\renewcommand{\theequation}{\thesection.\arabic{equation}}

\def\be{\begin{equation}}
\def\ee{\end{equation}}
\def\bea{\begin{eqnarray}}
\def\eea{\end{eqnarray}}
\def\m{\mu}
\def\n{\nu}
\def\g{\gamma}
\def\p{\phi}
\def\L{\Lambda}
\def \W{{\cal W}}
\def\bn{\bar{\nu}}
\def\bm{\bar{\mu}}
\def\bw{\bar{w}}
\def\ba{\bar{\alpha}}
\def\bb{\bar{\beta}}

\begin{titlepage}

\version\versionno

\hbox to \hsize{\hbox{\tt hep-th/0504155}\hss
	\hbox{\small{\tt MCTP-05-73}}}

\vspace{1.7cm}

\centerline{\bf \Large Towards Supergravity Duals of Chiral Symmetry
Breaking}

\vspace{.6cm}
\centerline{\bf \Large in Sasaki-Einstein Cascading Quiver Theories}
\vspace{2cm}

\centerline{ \large Benjamin  A. Burrington, James T. Liu}

\vspace{0.3cm}

\centerline{\large Manavendra
Mahato and Leopoldo A. Pando Zayas}

\vspace{1cm}

\centerline{\it Michigan Center for Theoretical Physics}
\centerline{\it Randall Laboratory of Physics, The University of Michigan}
\centerline{\it Ann Arbor, MI 48109--1040}

\vspace{1cm}

\begin{abstract}
We construct a first order deformation of the complex structure of the cone over
Sasaki-Einstein spaces $Y^{p,q}$ and check supersymmetry
explicitly. This space is a central element in the holographic dual of
chiral symmetry breaking for a large class of cascading quiver
theories. We discuss a solution describing a stack of $N$ D3 branes
and $M$ fractional D3 branes at the tip of the deformed spaces.
\end{abstract}

\end{titlepage}

\section{Introduction and summary}

One of the most promising elements of the AdS/CFT correspondence is
the possibility of providing an alternative method of studying aspects
of confining theories. Unfortunately, there are but a few smooth supergravity
backgrounds potentially dual to confining ${\cal N}=1$ SYM.

Recently, a new class of Sasaki-Einstein manifolds $Y^{p,q}$ has been constructed
\cite{se,GMSW}. Given a Sasaki-Einstein manifold $X^5$ one can
consider a stack of $N$ D3 branes at the tip of the cone over
$X^5$. Taking the Maldacena limit leads to a duality between string
theory on $AdS_5\times X^5$ and a superconformal gauge theory living
in the world volume of the D3 branes. In this context the infinite family of
spaces $Y^{p,q}$ was shown to be dual to superconformal quiver gauge 
theories \cite{MS,sequiver}. Remarkably, the correspondence has
provided a better treatment of some of the gauge theory questions. 
In particular, the irrational nature of some R-charges has
been elucidated in the field theory using gravity input \cite{MS,friends}.

Since $Y^{p,q}$ spaces are generalizations of $T^{1,1}$ one naturally
wonders about the possibility of further generalizing these spaces
in a way that parallels the program carried by Klebanov and
collaborators for the conifold \cite{igor,KS}. In fact, the first step
in this direction 
has already
been taken in a recent collaboration including Klebanov \cite{HEK}. The pinnacle of
Klebanov's program is the Klebanov-Strassler (KS) background \cite{KS}, which is a
smooth supergravity solution constructed as a warped deformed
conifold.

In this paper we take a step in the construction of
generalizations of the KS background. Namely, we construct a first
order deformation of the complex structure of the cone over $Y^{p,q}$.
This deformation should lead, upon
warping, to a supergravity background describing the IR of the
recently constructed cascading solution \cite{HEK} on the cone over $Y^{p,q}$.

One important motivation for this work is that the existence of a
supergravity background allows for explicit computation of
some dynamical quantities, like the spectrum of various states
including mesons, quantum corrections to Regge trajectories and possibly Goldstone
bosons.

In section \ref{review} we review some of the key aspects of the
$Y^{p,q}$ spaces and their dual quiver gauge theories. Our aim is to
set up the notation and technical motivation for later sections. In
particular, we address various technical aspects as the supergravity
realization of chiral symmetry breaking and the algebra of vielbeins.
In section \ref{deformation}, after recalling the general theory of deformation
of CY spaces from the differential geometric point of view, we
present the first order deformation for the cone over $Y^{p,q}$. This
is the main technical result of the paper. Section \ref{deformation} also
contains the result of explicitly checking that the deformation is
supersymmetric. This guarantees that the deformation has $SU(3)$
holonomy and that indeed corresponds to a complex K\"ahler manifold. Section
\ref{kah} reviews the complex coordinates introduced for the cone over
$Y^{p,q}$ in \cite{MS} with the aim of classifying the deformation. We
show that it corresponds to a deformation of the complex
structure. Moreover, in this section we find the K\"ahler potential
for the cone over $Y^{p,q}$, knowing the K\"ahler potential opens a
new venue for understanding deformations of these CY spaces. Section
\ref{warp} discusses placing a stack of $N$ D3 and $M$ fractional D3 branes at the tip of the
deformed space. Since we only know the deformation of the space to
first order, our analysis is approximate. The main technical result in
this section is the explicit construction of the imaginary self dual
3-form in the deformed space. This implies that to first order there
is a supergravity background that should correspond to the chiral
symmetry broken phase of the cascading Sasaki-Einstein quivers
discussed in \cite{HEK}.

Let us end this section with some comments on open questions that we
were not able to answer in this work. The most glaring question is the
existence of a solution beyond first order. 
Our paper follows a strictly differential geometric approach
to the deformation. Another unexplored venue is the algebraic
approach. Namely, the complex deformation for the conifold is fairly
simple from the algebraic standpoint. Given the
defining equation $\sum\limits_{a=1}^4 w_a^2=0$, we simply need to
replace it by $\sum\limits_{a=1}^4 w_a^2=\epsilon^2$. There are
various questions that the algebraic approach answers immediately. For
example, chiral symmetry breaking is nothing but the breaking of the
$U(1)$ symmetry that rotates the coordinates $w_a$ in the undeformed
space. In this paper we followed exclusively the differential
geometric approach but we hope to discuss the algebraic one in the
future. Finally, chiral symmetry breaking can also be explored along
the lines of \cite{chiral,chiral2}.

\section*{Note Added}
After the first version of this work appeared in the archive, a series of
papers addressing the complex deformation from the field theory side
have given arguments in favor of supersymmetry breaking
\cite{susybreaking}. 
We also became aware of work by Altmann \cite{altmann}
who shows the existence of an obtruction for finding the complex
deformations beyond first order. It is worth mentioning that while
these works make a case for obstruction to a solution built around a
complex deformation of the corresponding cone over $Y^{p,q}$ it is
plausible  that the full solution  is supersymmetric albeit with
$SU(3)$-structure rather than $SU(3)$ holonomy as is the case for the
baryonic branch of the KS solutions as discussed in \cite{minasian}.

\section{Review of superconformal quiver theories and their dual
$AdS_5\times Y^{p,q}$ backgrounds}\label{review}

The gauge theory dual to IIB on $AdS_5\times Y^{p,q}$ has been the
subject of much recent investigation. Here we begin with a summary
of some of the key aspects of the quiver gauge theories with $Y^{p,q}$
duals, as explained in Ref.~\cite{sequiver}.

The quivers for $Y^{p,q}$ can be constructed starting with the quiver
of $Y^{p,p}$ which is naturally related to the quiver theory obtained
from $\mathbb C^3/\mathbb Z_{2p}$. The gauge group is $SU(N)^{2p}$ and the
superpotential is constructed out of cubic and quartic terms in the
four types of fields present: $U,V,Y$ and $Z$, where $U$ and $V$ are
doublets of $SU(2)$:
\be
\epsilon_{\a\b} U^\a_a V^\b Y, \qquad \epsilon_{\a\b} U^\a_b V^\b Y,
\qquad \epsilon_{\a\b}ZU_a^\a Y U_b^\b.
\ee
Greek indices $\a,\b=1,2$ are in $SU(2)$, and Latin subindices $a,b$
refer to the gauge group where the corresponding arrow originates.
Equivalently, as explained in \cite{HEK}, the quiver theory for $Y^{p,q}$
can be constructed from two basic cells denoted by $\sigma$ and $\tau$.
Some concrete examples can be found in \cite{sequiver,HEK}.

One key ingredient is that the geometric realization of the
${\bf R}$-symmetry is given by a Killing vector field in the
Sasaki-Einstein geometry of the form
\be
\eqlabel{u1inse}
K=\frac13 \frac{\partial}{\partial \psi},
\ee
where the geometrical meaning of the coordinates $\psi$ and $\alpha$
will become apparent momentarily. Breaking the ${\bf R}$-symmetry is
our main goal, and it translates into constructing a supergravity
background without (\ref{u1inse}) as a Killing vector.

\subsection{The cone over $Y^{p,q}$}

Before turning on any deformations, we highlight the structure of the
cone over $Y^{p,q}$, emphasizing its similarities with the conifold,
which is the cone over $T^{1,1}$.  This connection motivates our later
Ansatz for the deformation.

Following \cite{GMSW,MS}, the Sasaki-Einstein metric on $Y^{p,q}$
can be written in the form
\begin{eqnarray}
ds^2&=&\frac{1-cy}{6}(d{\theta}^2+{\sin}^2 \theta d{\phi}^2)
+\frac{1}{w(y)v(y)}dy^2
+\frac{w(y)v(y)}{36}(d\beta-c\cos\theta d\phi )^2\nonumber\\
&&+\frac{1}{9}[d\psi +\cos\theta d \phi+y(d\beta-c \cos\theta d\phi )]^2\\
&=&\frac{1}{6}[(e^{\theta})^2+(e^{\phi})^2+(e^y)^2+(e^{\beta})^2]
+\frac{1}{9}(e^{\psi})^2,
\end{eqnarray}
where
\begin{equation}
w(y)=\frac{2(a-y^2)}{1-cy}, \qquad v(y)=\frac{a-3y^2+2cy^3}{a-y^2}.
\end{equation}
The natural one forms are given by \cite{HEK,SSP}
\begin{eqnarray}
e^{\theta}&=&\sqrt{1-cy}\;d\theta, \nonumber\\
e^{\phi}&=&-\sqrt{1-cy}\;\sin\theta d\phi, \nonumber\\
e^y&=&-\frac{1}{H(y)}dy, \nonumber\\
e^{\beta}&=& H(y)(d\beta-c \cos\theta d\phi), \nonumber\\
e^{\psi}&=& d\psi+\cos\theta d\phi+y(d\beta-c\cos\theta d\phi),
\end{eqnarray}
with $H(y)=\sqrt{w(y)v(y)/6}$.

Before proceeding, it would be convenient to develop the algebra of
exterior derivatives. For this purpose, let
$L(y,\theta)=\frac{\cot\theta}{\sqrt{1-cy}}$
and $K(y)=\frac{cH(y)}{2(1-cy)}$. Then $\frac{dH(y)}{dy}=K(y)-\frac{y}{H(y)}$.
The exterior derivatives of the above forms then satisfy
\begin{eqnarray}
de^{\theta}&=& K(y)e^{y}\wedge e^{\theta}, \nonumber\\
de^{\phi}&=&K(y)e^y\wedge e^{\phi}+L(y,\theta)e^{\theta}\wedge
e^{\phi}, \nonumber\\
de^{y}&=&0, \nonumber\\
de^{\beta}&=&\left (\frac{y}{H(y)}-K(y)\right )
e^y\wedge e^{\beta}-2K(y)e^{\theta}\wedge e^{\phi}, \nonumber\\
\label{de}
de^{\psi}&=&e^{\theta}\wedge e^{\phi}-e^y\wedge e^{\beta}.
\end{eqnarray}

At various stages in the construction, we will refer to the model of
the deformed conifold. In particular, it is worth noting that the
above definitions reduce to those in \cite{KS} describing $T^{1,1}$ in
the limit \cite{GMSW}
\begin{equation}
\eqlabel{limit}
c\rightarrow 0,\qquad
y\rightarrow \cos\theta _2,\qquad
a \rightarrow 3 \quad\hbox{and}\quad
\beta \rightarrow \phi_2.
\end{equation}
Following the discussion of \cite{KS} (see also \cite{mt}), one can further define
the rotated and shifted vielbeins $e^1,e^2,\ldots,e^5$ and
$g^1,g^2,\ldots,g^5$ as
\begin{eqnarray}
e^1=-e^{\beta},&&
e^3=-\cos\psi e^{\phi}-\sin\psi e^{\theta},\qquad e^5=e^{\psi},\nonumber\\
e^2=e^{y},\hphantom{-}&&
e^4=-\sin\psi e^{\phi}+\cos\psi e^{\theta},
\label{eq:ei}
\end{eqnarray}
and
\begin{eqnarray}
g^1=\frac{1}{\sqrt{2}}(e^1-e^3),&&
g^3=\frac{1}{\sqrt{2}}(e^1+e^3),\qquad g^5=e^{5}.\nonumber\\
g^2=\frac{1}{\sqrt{2}}(e^2-e^4),&&
g^4=\frac{1}{\sqrt{2}}(e^2+e^4),
\label{gs}
\end{eqnarray}

Note that, if desired, one could instead have defined
$\tilde{e}^1,\tilde{e}^2,\ldots,\tilde{e}^5$ as
\begin{eqnarray}
\tilde{e}^1=e^{\phi},&&
\tilde{e}^3=\cos\psi e^{\beta}-\sin\psi e^y,\qquad \tilde{e}^5=e^{\psi}.
\nonumber\\
\tilde{e}^2=e^{\theta},&&
\tilde{e}^4=\sin\psi e^{\beta}+\cos\psi e^y,
\label{eq:other}
\end{eqnarray}
The corresponding $\tilde{g}^1,\tilde{g}^2,...,\tilde{g}^5$ would then be
defined in a similar fashion as in (\ref{gs}) with $e^i$ replaced by
$\tilde{e}^i$ for $i=1,2,\ldots,5$.  One should note, however, that
the $\tilde{e}^i$ are physically indistinct from the ${e}^i$ in
the $T^{1,1}$ limit (\ref{limit}).  They are distinct when $c\neq0$, and so
we may expect that they could play a r\^ole in deforming the cone over
$Y^{p,q}$.  We will nevertheless show that the first order deformation
given in terms of the $\tilde e^i$ is no different than that given in
terms of the $e^i$.

Although the angular coordinate $\psi$ appears explicitly in the the
above one forms, the metric remains $\psi$ independent.  As we will see
in the next section, the perturbed solution will break this symmetry
direction in an analogous way to the solution of \cite{KS}.

\section{The first order complex deformation}\label{deformation}

We now briefly review the deformation theory of CY spaces following
the presentation of \cite{moduli}. Let the parameter space
of Calabi-Yau manifolds be the parameter space of Ricci-flat
K\"ahler metrics, and let $g_{mn}$ and $g_{mn}+ \delta g_{mn}$
satisfy
\be
R_{mn}(g)=0, \qquad R_{mn}(g+\delta g)=0.
\ee
Then, with the coordinate condition $\nabla^n\delta g_{mn}=0$, one
obtains that $\delta g_{mn}$ satisfies the Lichnerowicz equation
\be
\eqlabel{lich}
\nabla^k\nabla_k \delta g_{mn}+ 2R_m{}^p{}_n{}^q \delta g_{pq}=0.
\ee

The connection between the geometro-differential approach and the
algebraic approach to deforming a CY space arises due to an
isomorphism between the solutions of (\ref{lich}) and harmonic forms on
CY. Namely, a solution with mixed indices is associated with a
$(1,1)$-form
\be
i\delta g_{\m\bar{\nu}}dx^\m \wedge dx^{\bar{\nu}},
\ee
which is harmonic if and only if the variation of the metric satisfies
the Lichnerowicz equation, (\ref{lich}). Similarly, for a variation of
pure type, one can associate a $(2,1)$-form using the holomorphic 3-form
$\Omega$:
\be
\Omega_{\kappa \lambda}{}^{\bar{\nu}} \delta g_{\bar{\mu}\bar{\nu}}
dx^{\kappa}\wedge dx^{\lambda}\wedge dx^{\bar{\mu}}.
\ee
This form is harmonic if and only if $\delta g_{\bar{\mu}\bar{\nu}}$
satisfies (\ref{lich}).

With these isomorphisms in place we can classify a deformation as either
K\"ahler or complex.  In particular, variations of pure type (there are
$b_{2,1}$ of them) correspond to variations of the complex structure. Note
that $g+\delta g$ is a K\"ahler metric on a manifold close to the original
one.  There must therefore exist a coordinate system in which the pure parts
of the metric vanish. Under a change of coordinates $x^m \to x^m +
f^m(x)$, the metric variation transforms according to the familiar
\be
\delta g_{mn}\to \delta g_{mn} - \frac{\partial f^r}{\partial x^m }
g_{rn} - \frac{\partial f^r}{\partial x^n } g_{mr}.
\ee
If $f^\mu$ is holomorphic then $\delta g_{\m\n}$ is invariant. Thus,
the pure part of the variation could be removed by a transformation of
coordinates but it cannot be removed by a holomorphic coordinate
transformation. Thus the pure part of the metric variation corresponds
precisely to changes of the complex structure.

\subsection{First order perturbation}

We are now in a position to construct the first order deformation of the
cone over $Y^{p,q}$.  Using the one-forms of section \ref{review}, the
undeformed metric is given by
\be
ds^2 = dr^2+r^2\left(\ft16\left[(e^1)^2+(e^2)^2+(e^3)^2+(e^4)^2\right]
+\ft19(e^5)^2\right).
\label{eq:cypq}
\ee
This has the same form as the conifold; however important nontrivial
dependences on the coordinate $y$ have been introduced through the
definitions of the vielbeins $e^i$.

As mentioned above, most of our intuition for constructing a deformation
of (\ref{eq:cypq}) comes from the analogous case of the deformed conifold.
Here we recall that the deformed conifold metric can be written as
follows \cite{conifold}:
\bea
ds_{DC}^2= \epsilon^{\fft43}\frac{K(\tau)}{3}\bigg(\frac{1}{3K^3}
\left[4\,d\tau^2+(g^5)^2\right] \!\!&+&\!\!
\cosh(\tau)^2\left[(g^3)^2+(g^4)^2\right]\nonumber \\
\!\!&+&\!\! \sinh(\tau)^2 \left[(g^1)^2+(g^2)^2\right]\bigg),
\eea
where
\be
K(\tau) = \frac{ 2^{2/3}(\sinh(4\tau)-4\tau)^{1/3}}{2 \sinh(2\tau)}.
\ee

While the above is given in terms of a radial coordinate $\tau$,
a more natural coordinate for our analysis is given by $r$ where
$r^3=\epsilon^2 \cosh(2 \tau)$.  Furthermore, to understand the first order
deformation, is it also convenient to expand the metric for large $r$;
more precisely, we expand in the dimensionless quantity $\epsilon^2/r^3$.
We find, up to second order in the deformation, that
\bea
ds_{DC(2)}^2 &=&
dr^2+ r^2\left(\fft{1}{6}\left[(g^1)^2+(g^2)^2+(g^3)^2+(g^4)^2\right]
+\fft{1}{9} (g^5)^2\right) \nonumber \\
&&+\fft16r^2\left[-(g^1)^2-(g^2)^2+(g^3)^2+(g^4)^2\right]
\frac{\epsilon^2}{r^3}\nonumber \\
&&+ \frac{1}{3}\bigg(-\frac{1}{6} r^2\bigl[(g^3)^2+(g^4)^2+(g^1)^2+(g^2)^2
\bigr] \nonumber \\
&& \qquad\qquad+2\left[dr^2+\frac{1}{9}(g^5)^2r^2\right]\bigg)
\left(\ln\left(\frac{2r^3}{\epsilon^2}\right)-1\right)
\frac{\epsilon^4}{r^6}+dr^2\frac{\epsilon^4}{r^6}+\cdots. \nonumber\\
\label{eq:dc2}
\eea
The zeroth order term is of course the undeformed conifold metric itself.
The first order (in $\epsilon^2$) term should be a solution to the
linearized Einstein's equations, and so should be a zero mode of the
Lichnerowicz operator (\ref{lich}).  We make the simple observation that
the first order contribution is transverse and traceless.  The second
order perturbation is ``sourced" by the first order equations, and we
have verified explicitly that the above system indeed satisfies the
Ricci-flatness condition to second order.

We now turn our attention back to the cone over $Y^{p,q}$.  We note that
the combination $-(g^1)^2-(g^2)^2+(g^3)^2+(g^4)^2$ showing up at first
order in the above expansion is the same as $2(e^1e^3+e^2e^4)$.  This
therefore suggests an Ansatz for a first order perturbation for the cone
over $Y^{p,q}$ of the form
\be
ds_{(1)}^2= h_{m n} dx^m dx^n = f(y)r^2\frac{\left(e^1e^3+e^2e^4\right)}{3}
\frac{\epsilon^2}{r^3},
\label{eq:firsto}
\ee
where we have included a function of $y$ because of the non trivial
dependence of the $Y^{p,q}$ metric on this coordinate.  Inserting this
Ansatz into the first order deformation equations, we find that $f(y)$
must satisfy a {\it first order ODE} for it to be a zero mode (even though the
Lichnerowicz operator is {\it second} order).  The solution is simply
\be
f(y)=\frac{1}{(1-cy)^2}.
\ee
We have dropped an arbitrary multiplicative factor (it only becomes
important in the case of a fully non-perturbative solution).  Note
that the first order perturbation is again transverse and traceless.

Despite some effort, we have unfortunately not been able to satisfy the second
order deformation equations using the first order contributions as a source.
It is likely that, at second order, the deformation of the metric will
non-trivially depend on both $r$ and $y$ coordinates in an inherently
non-separable manner.  Although we have no direct proof, this belief
is supported by various unsuccessful attempts at separating the functional
dependence on $r$ and $y$ in the spirit of the deformed conifold metric
(\ref{eq:dc2}).

When constructing the one-forms $e^i$, we have introduced angular
$\psi$ dependence by ``rotating" $e^{\phi}$ and $e^{\theta}$ together
using $\psi$.  As mentioned above, one could instead have mixed
$e^{y}$ and $e^{\beta}$.  This, however, does not alter the first order
perturbation (up to irrelevant minus signs).  Hence the other mixing
(\ref{eq:other}), while it is perhaps distinct non-perturbatively, is
indistinct to lowest order in $\epsilon^2$. This might signal another
possible type of deformation, but we will not explore it here.

As mentioned above, the rotation between $e^{\phi}$ and $e^{\theta}$
introduces dependence on the coordinate $\psi$.  This dependence on
$\psi$ disappears in the zeroth order metric because $e^{\phi}$ and
$e^{\theta}$ always appear in the $SU(2)$ invariant combination.  However
this is no longer true for the first order perturbation.  This dependence
on $\psi$ is dual to chiral symmetry breaking in the gauge theory, and
will be discussed more explicitly in section \ref{pured3} below.

\subsection{Supersymmetry of the perturbed solution}\label{secSUSY}

Although the first order deformation was simply obtained by demanding
Ricci-flatness to order $\epsilon^2$, it turns out that it is in fact a
complex deformation, at least to this order.  To show this explicitly,
we turn to the Killing spinor equation.  In the absence of fluxes, the
supersymmetry condition takes the form $\nabla \lambda=0$, with the
resulting $\lambda$ a parallel spinor.

However, instead of working with the Killing spinor equation directly,
we focus on the integrability condition
\be
\eqlabel{riemProj}
e_a^\mu e_b^\nu\left[\nabla_\mu, \nabla_\nu \right]\lambda
=\ft14 R_{abcd} \Gamma^{cd} \lambda=0.
\ee
Before obtaining the Riemann tensor in the tangent basis, we first make
a convenient choice of vielbein.  To do this, we note that we can
absorb the first order term (\ref{eq:firsto}) into a shift of $e^1$
and $e^2$ according to
\be
\eqlabel{1order}
ds^2 = dr^2+r^2\left(\ft16 \left[(e^1+\delta e^3)^2+(e^2+\delta e^4)^2
+ (e^3)^2 + (e^4)^2\right]+\ft19 (e^5)^2\right),
\ee
where we have defined the quantity
\be
\eqlabel{delta}
\delta = \frac{\epsilon^2}{r^3(1-cy)^2}.
\ee
This introduces a shift to second order in $\epsilon^2$, which however
is unimportant as we will always work only to first order in $\epsilon^2$.
We now make a natural choice of shifted vielbeins
\bea
\hat{e}^6_{(s)}= dr,~&&~
\hat{e}^1_{(s)}=\frac{r}{\sqrt{6}}\left(e^1+\delta e^3\right),
\qquad \hat{e}^2_{(s)}=\frac{r}{\sqrt{6}}\left(e^2+\delta e^4\right),
\nonumber\\
\hat{e}^3_{(s)}=\frac{r}{\sqrt{6}}e^3,~&&~
\hat{e}^4_{(s)}=\frac{r}{\sqrt{6}}e^4,
\kern5.8em \hat{e}^5_{(s)}=\frac{r}{3}e^5.
\label{eq:ncsv}
\eea

Using this shifted vielbein basis, we find that the integrability
condition (\ref{riemProj}) is satisfied for spinors $\lambda$ satisfying
the simultaneous projections
\bea
\left(1+\Gamma^{1256}\right)\lambda &=& 0, \nonumber \\
\left(1-\Gamma^{1234}\right)\lambda &=& 0.
\eea
Writing out the $SO(6)$ generators in the spinor representation as
$T^1=\fft{i}2\Gamma^{12}$, $T^2=\fft{i}2\Gamma^{34}$ and
$T^3=\fft{i}2\Gamma^{56}$, we see that parallel spinors have $SO(6)$
weights $(\fft12,-\fft12,\fft12)$ or $(-\fft12,\fft12,-\fft12)$.
Changing to an $SU(4)$ basis with Cartan generators
\begin{equation}
H^1=\ft12(T^1+T^2),\qquad H^2=\ft1{4\sqrt{3}}(-T^1+T^2+2T^3),\qquad
H^3=\ft1{2\sqrt{6}}(T^1-T^2+T^3),
\end{equation}
we see that the parallel spinors are singlets of the $SU(3)$ corresponding
to $H^1$ and $H^2$.  This is just the standard Calabi-Yau decomposition
$\mathbf4\to\mathbf3+\mathbf1$ under $SU(4)\supset SU(3)$.

More explicitly, we have verified that the integrability condition
(\ref{riemProj}) is satisfied at both zeroth and first order in $\epsilon^2$
for the above projections.  For example%
\footnote{All components have been computed in Maple.  Details of the
calculations are available upon request to the authors.}
\bea
R_{53 ab}\Gamma^{ab} &=& \frac{6\epsilon^2}{r^5(1-cy)^2}
\Gamma^{51}\left(1+\Gamma^{1256}\right),\nonumber\\
R_{24 ab}\Gamma^{ab} &=& \frac{2(ac^2-1)}{r^2(1-cy)^3}\Gamma^{31}
\left(1-\Gamma^{1234}\right)
+\frac{3\sqrt{6}H(y)\epsilon^2 c}{r^5(1-cy)^3}\Gamma^{51}
\left(1+\Gamma^{1256}\right),
\eea
which are clearly zero when applied to $\lambda$.

\section{Complex coordinates and K\"ahler potential}\label{kah}

Having given the first order deformation and shown that it is
supersymmetric, we now proceed to its classification. As a first step,
we review the complex coordinates introduced in \cite{MS} with some minor
adjustments to fit the present conventions%
\footnote{Note that we have taken $\phi \to -\phi$ with respect
to \cite{MS}.}.
The main new result of this section is an expression for the K\"ahler
potential of the cone over $Y^{p,q}$.

We recall that Ref.~\cite{MS} obtained the complex coordinates for
the cone over Sasaki-Einstein space. Let
\begin{eqnarray}
{\eta}^1&=&\frac{1}{\sin\theta}d\theta-id\phi, \nonumber\\
\tilde{\eta}^2&=&-\frac{dy}{H(y)^2}-i(d\beta-c \cos\theta d\phi), \nonumber\\
\tilde{\eta}^3&=&\frac{3dr}{r}+i\big[d\psi+\cos\theta d\phi+y(d\beta-c
\cos\theta d\phi)\big].
\end{eqnarray}
In terms of these coordinates, one can write the metric as
\begin{equation}
ds^2=r^2\frac{(1-cy)}{6}{\sin}^2\theta {\eta}^1\overline{\eta^1}
+r^2\frac{H(y)^2}{6}{\tilde{\eta}^2}\overline{\tilde{\eta}^2}
+\frac{r^2}{9}{\tilde{\eta}^3}\overline{\tilde{\eta}^3}.
\end{equation}
Unfortunately, ${\tilde{\eta}^2}$ and ${\tilde{\eta}^3}$ are not
integrable. However, integrable one-forms can be obtained by taking linear
combinations of them:
\be
{\eta}^2={\tilde{\eta}^2}+c \cos \theta {\eta}^1, \qquad
{\eta}^3={\tilde{\eta}^3}+\cos\theta {\eta}^1+y{\tilde{\eta}^2}.
\ee
In this case, ${\eta}^i=dz_i/z_i$ for $i=1,2,3$, where
\begin{eqnarray}
\label{z1}
z_1&=&\tan\frac{\theta}{2}e^{-i\phi},\\
\label{z2}
z_2&=&\frac{(\sin\theta)^{2c}}{f_1(y)}e^{-i\beta},\\
\label{z3}
z_3&=&r^3\frac{\sin\theta}{f_2(y)}e^{i\psi},
\end{eqnarray}
with
\begin{equation}
f_1(y)=\exp\left(\int\frac{1}{H(y)^2}dy\right),\qquad
f_2(y)=\exp\left(\int\frac{y}{H(y)^2}dy\right).
\end{equation}
As a check, we may take the limit to the conifold (\ref{limit}), in
which case the $z_i$'s reduce to
\begin{equation}
z_1\rightarrow \tan\frac{{\theta}_1}{2}e^{-i{\phi}},\qquad
z_2\rightarrow \tan\frac{{\theta}_2}{2}e^{-i{\beta}},\qquad
z_3\rightarrow r^3\sin{\theta}_1\sin{\theta}_2e^{i\psi},
\end{equation}
along with $f_1=\cot(\theta_2/2), \quad f_2=\csc\theta_2$.

Returning to the Sasaki-Einstein case, note that
(\ref{z1}) can be written as
\begin{equation}
\label{t}
\sin\theta=\frac{2\sqrt{z_1\bar{z}_1}}{1+z_1\bar{z}_1},
\end{equation}
where bars denote complex conjugation.  Therefore
\begin{equation}
\label{dt}
\frac{\partial\sin\theta}{\partial
z_1}=\frac{1}{2z_1}\sin\theta\cos\theta,\qquad
\frac{\partial\sin\theta}{\partial z_2}=0,\qquad
\frac{\partial\sin\theta}{\partial z_3}=0.
\end{equation}
{}From equation (\ref{z2}), an explicit expression for $f_1(y)$ can be
obtained:
\begin{equation}
f_1(y)^2=\frac{({\sin\theta})^{2c}}{z_2\bar{z}_2}=\sigma.
\end{equation}
We are unable, however, to invert this equation to obtain an explicit
expression of $y$ in terms of $\sigma$.  Nevertheless, the above equation
suggests that if an explicit expression for $y$ is possible, then the
complex coordinates will enter into it only in the above combination
denoted by $\sigma$. Moreover, $y$ is independent of $z_3$ and its conjugate,
{\it i.e.} $\partial y/\partial z_3=\partial y/\partial \bar{z}_3=0$.

For suitable values of $y$, the total derivative of $y$ can be evaluated:
\begin{equation}
\frac{dy}{d\sigma}=(\frac{d\sigma}{dy})^{-1}
=\frac{wv}{12f_1(y)^2}.
\end{equation}
This is possible because $y$ can be viewed as a function of $\sigma$ only.
In turn, $\sigma$ is a function of $z_1,z_2$ and their conjugates. One
obtains
\begin{equation}
\frac{\partial \sigma}{\partial z_1}=\frac{c\sigma}{z_1}\cos\theta.
\end{equation}
Then
\begin{equation}
\frac{\partial y}{\partial z_1}=\frac{dy}{d\sigma}
\frac{\partial\sigma}{\partial z_1}=\frac{1}{z_1}\frac{wv}{12}c\cos\theta.
\end{equation}
Similarly, other partial derivatives can be evaluated. One can write
all of them more succinctly using coordinates $u_i=\ln z_i$ as
\begin{equation}
\begin{tabular}{llll}
$\displaystyle\frac{\partial \theta}{\partial u_1}=\frac{1}{2}\sin\theta,$
&$\displaystyle\frac{\partial \theta}{\partial u_2}=0,$
&$\displaystyle\frac{\partial \theta}{\partial u_3}=0,$
&$\displaystyle\frac{\partial \theta}{\partial \bar{u}_i}=
\frac{\partial \theta}{\partial u_i},$\\[2ex]
$\displaystyle\frac{\partial y}{\partial u_1}=\frac{wv}{12}c\cos\theta,$
&$\displaystyle\frac{\partial y}{\partial u_2}=-\frac{wv}{12},$\kern1em
&$\displaystyle\frac{\partial y}{\partial u_3}=0,$\kern1em
&$\displaystyle\frac{\partial y}{\partial \bar{u}_i}=
\frac{\partial y}{\partial u_i},$\\[2ex]
$\displaystyle\frac{\partial r}{\partial u_1}=-\frac{r}{6}(1-cy)\cos\theta,$
\kern1em
&$\displaystyle\frac{\partial r}{\partial u_2}=-\frac{ry}{6},$
&$\displaystyle\frac{\partial r}{\partial u_3}=\frac{r}{6},$
&$\displaystyle\frac{\partial r}{\partial \bar{u}_i}=
\frac{\partial r}{\partial u_i}.$\\
\end{tabular}
\end{equation}

The metric can now be written in terms of the $u_i$ as
\begin{eqnarray}
ds^2&=&\left [r^2\frac{(1-cy)}{6}{\sin}^2\theta
+r^2\frac{wv}{36}c^2{\cos}^2{\theta}
+\frac{r^2}{9}(1-cy)^2{\cos}^2\theta
\right ]du_1 d\bar{u}_1\nonumber\\
&&+\left [ r^2\frac{wv}{36}+\frac{r^2}{9}y^2\right ]du_2 d\bar{u}_2
+\frac{r^2}{9}du_3 d\bar{u}_3\nonumber\\
&&+\left [-r^2\frac{wv}{36}c\cos\theta+\frac{r^2}{9}y(1-cy)\cos\theta\right]
(du_1 d\bar{u}_2+d\bar{u}_1 du_2)\nonumber\\
\label{met}
&&-\frac{r^2}{9}y(du_2 d\bar{u}_3+ d\bar{u}_2 du_3)
-\frac{r^2}{9}(1-cy)\cos\theta (du_1 d\bar{u}_3+d\bar{u}_1 d{u_3}).
\end{eqnarray}
The coordinate $r$ in the limit (\ref{limit}) reduces to the coordinate
$\rho$ as used by \cite{conifold} for the conifold. Recall that the K\"ahler
potential in the conifold case is ${\rho}^2$. Here, one can show that the
K\"ahler potential for the Sasaki-Einstein cone is $r^2$.  For example,
\begin{eqnarray}
{\partial}_{u_1}{\partial}_{\bar{u}_1}r^2&=&
{\partial}_{u_1}\left(-2r\frac{r}{6}(1-cy)\cos\theta \right )\nonumber\\
&=&-\left(\frac{1}{3}\frac{{\partial}r^2}{\partial{u_1}}(1-cy)\cos\theta
+\frac{r^2}{3}\frac{{\partial}(-cy)}{\partial{u_1}}\cos\theta
+\frac{r^2}{3}(1-cy)\frac{{\partial}\cos\theta}{\partial{u_1}}\right)
\nonumber\\
&=&\frac{r^2}{9}(1-cy)^2{\cos}^2\theta
+r^2\frac{wv}{36}c^2{\cos}^2\theta
+\frac{r^2}{6}(1-cy){\sin}^2{\theta},
\end{eqnarray}
which matches with the $(1,\bar{1})$ component of the metric written in
(\ref{met}).  The other components of the metric can be similarly
obtained. Thus the relation $g_{{\mu}\bar{\nu}}
={\partial}_{\mu}{\partial}_{\bar{\nu}}{\cal K}$ holds for the K\"ahler
potential ${\cal K}=r^2$. This is an interesting result, and due to its
simplicity it could provide alternative ways of understanding
deformations and resolutions 
of the cone over $Y^{p,q}$. Similarly, the K\"ahler potential
for the four-dimensional K\"ahler-Einstein base turns out to be
$\frac{2}{3}\ln[(1+\frac{1}{z_1\bar{z}_1}) f_2(y)]$.

\subsection{First order perturbation in complex form}

We now arrive at our goal of showing that the deformation obtained in
the previous section is a complex deformation.  Using the complex one-forms
defined earlier, one obtains
\begin{eqnarray}
-\sin\theta{\eta}^1\tilde{\eta}^2&=&
\frac{1}{H(y)^2}dyd\theta+\sin\theta d\phi (d\beta- c\cos\theta d\phi )
\nonumber\\&&
+i\big[d\theta (d\beta-c\cos\theta d\phi )
-\frac{1}{H(y)^2}dy\sin\theta d\phi \big].
\end{eqnarray}
In terms of these complex coordinates, the metric perturbation can be written as
\begin{eqnarray}
\label{gcomplex}
2(e^1e^3+e^2e^4)
&=&2[\cos{\psi}(e^{\beta}e^{\phi}+e^{y}e^{\theta})
+\sin{\psi}(e^{\theta}e^{\beta}-e^ye^{\phi})]\nonumber\\
&=&-2H(y)\sqrt{1-cy}\;[
\cos\psi \,{\rm Re}(-\sin\theta {\eta}^1\tilde{\eta}^2 )
+\sin\psi \,{\rm Im}(-\sin\theta {\eta}^1\tilde{\eta}^2)]\nonumber\\
\label{f-o}
&=&H(y)\sqrt{1-cy}\;\sin\theta [e^{i\psi}{\eta}^1\tilde{\eta}^2
+e^{-i\psi}\bar{\eta}^1\bar{\tilde{\eta}}^2].
\end{eqnarray}
The first order perturbation is of pure type.  We note here that
the $\eta^1 \tilde{\eta}^2$ part is {\it by itself} a zero mode of the
Lichnerowicz operator.  Since the holomorphic (3,0)-form for the
Sasaki-Einstein cone is known \cite{MS}, a complex closed $h_{2,1}$
form can be constructed,
\begin{eqnarray}
h_{(2,1)}&=&-\frac{1}{18}\frac{{\epsilon}^2}{(1-cy)^2}
[H^2du^2\wedge du^3\wedge d\bar{u}^2+{\sin}^2\theta (1-cy)
du^3\wedge du^1\wedge d\bar{u}^1\nonumber\\&&
+cH^2{\cos}\theta du^3\wedge du^1\wedge d\bar{u}^2
+y(1-cy){\sin}^2\theta du^1\wedge du^2\wedge d\bar{u}^1\nonumber\\&&
+H^2\cos\theta du^1\wedge du^2\wedge d\bar{u}^2].
\end{eqnarray}
%

\section{Warping the deformation}\label{warp}

The general form of the solution we are seeking in this section has
been explained in \cite{HEK}. The main difference is that instead
of warping the cone over $Y^{p,q}$ we need to warp the deformation of the
cone. The solution contains a nontrivial $F_5$ representing the flux
left by the D3 branes after the transition and $G_3$ which is the
implication of considering fractional D3 branes, that is, D5 branes
wrapping a 2-cycle inside the deformed cone. Thus the full IIB solution
is given in terms of the fields
\bea
\label{ansatz}
ds^2&=& h^{-1/2} \eta_{\mu\nu} dx^\mu dx^\nu + h^{1/2}ds_6^2,\nonumber\\
ds_6^2&=& dr^2 +r^2\left[(e^1_{(s)})^2+(e^2_{(s)})^2+(e^3_{(s)})^2+
(e^4_{(s)})^2+(e^5_{(s)})^2\right],\nonumber\\
F_5&=&(1+*)\calf_5=(1+*)dh^{-1}\wedge dx^0\wedge dx^1\wedge dx^2\wedge dx^3,
\nonumber\\
G_3&=&-F_3+\frac{i}{g_s}H_3= i M\Omega_{3}.
\eea
The shifted vielbein basis is given above in section \ref{secSUSY}.
Given this general form of the solution, our goal is to find the
explicit form of $h(r,y)$ and $ \Omega_{3}$. Furthermore, in the case
when the deformation is zero, we expect to recover the solution of \cite{HEK}.

\subsection{Turning on a $G_3$ flux in the deformed cone over $Y^{p,q}$}

In this section we search for an appropriate complex $G_3$ which may be
turned on in the deformed Sasaki-Einstein cone. To have a supersymmetric
solution we would need $G_3$ to be a $(2,1)$ form \cite{gubser}. However,
we will not address the Dolbeault decomposition of $G_3$ directly. Rather,
we will simply concentrate on finding a solution to the equations of motion
with constant dilaton field. The constancy of the dilaton is readily
satisfied by an imaginary self dual complex 3-form $G_3$.  We thus
consider $G_3$ to be proportional to an imaginary self dual 3-form
$\Omega_3$, namely
\be
\eqlabel{iself}
*_6\Omega_3=i\Omega_{3}.
\ee
To assist in taking the Hodge dual while at the same time keeping the
radial coordinate dependence explicit, we use a set of shifted vielbeins
\bea
e^6_{(s)}= dr, ~&&~ e^1_{(s)}=\left(e^1+\delta e^3\right),\qquad
e^2_{(s)}=\left(e^2+\delta e^4\right),\nonumber\\
e^3_{(s)}=e^3, ~&&~ e^4_{(s)}=e^4,
\kern5.7em e^5_{(s)}=e^5,
\eea
which resemble the choice of (\ref{eq:ncsv}), but with $r$-dependence
(and some factors) removed.

In this shifted vielbein basis, the most general Ansatz for an imaginary self
dual $\Omega_{3}$ may be written as
\bea
\Omega_{(3)}&=&\left(\frac{dr}{r}+ i\frac{e^5}{3}\right)\wedge
\bigg(\alpha_1\left(e^1_{(s)}\wedge e^2_{(s)}+e^3_{(s)}\wedge e^4_{(s)}\right)
+\alpha_2\left(e^1_{(s)}\wedge e^4_{(s)}+e^2_{(s)}\wedge e^3_{(s)}\right)
\nonumber \\
&&\kern8em+i\alpha_3\left(e^1_{(s)}\wedge e^3_{(s)}
-e^2_{(s)}\wedge e^4_{(s)}\right)\bigg) \nonumber \\
&&+\left(\frac{e^5}{3}+i\frac{dr}{r}\right)\wedge \bigg(
i\beta_1\left(e^1_{(s)}\wedge e^2_{(s)} - e^3_{(s)}\wedge e^4_{(s)}\right)
+i\beta_2\left(e^1_{(s)}\wedge e^4_{(s)} - e^2_{(s)}\wedge e^3_{(s)}\right)
\nonumber \\
&&\hphantom{+}\kern8em+\beta_3\left(e^1_{(s)}\wedge e^3_{(s)}
+e^2_{(s)}\wedge e^4_{(s)}\right)\bigg)\nonumber\\
&&+i\lambda_1\bigg(\frac{dr}{r} \wedge \frac{e_{(s)}^5}{3}\wedge e_{(s)}^1
+\frac{i}{6}e^2_{(s)}\wedge e^3_{(s)}\wedge e^4_{(s)}\bigg) \nonumber\\
&&+\lambda_2\bigg(\frac{dr}{r} \wedge \frac{e_{(s)}^5}{3}\wedge e_{(s)}^2
-\frac{i}{6}e^1_{(s)}\wedge e^3_{(s)}\wedge e^4_{(s)}\bigg) \nonumber\\
&&+i\lambda_3\bigg(\frac{dr}{r} \wedge \frac{e_{(s)}^5}{3}\wedge e_{(s)}^3
+\frac{i}{6}e^1_{(s)}\wedge e^2_{(s)}\wedge e^4_{(s)}\bigg) \nonumber\\
&&+\lambda_4\bigg(\frac{dr}{r} \wedge \frac{e_{(s)}^5}{3}\wedge e_{(s)}^4
-\frac{i}{6}e^1_{(s)}\wedge e^2_{(s)}\wedge e^3_{(s)}\bigg),
\label{eq:omeg3}
\eea
where the ten functions $\alpha_i,\beta_i,\lambda_i$ are general (possibly
complex) functions.  The factors of $i$ in front of some of these functions
are chosen for later convenience.  We, however, restrict to the case when
these functions depend only on $r$ and $y$.

Our goal now is to expand the above equations in $\delta$, that is in powers
of $\epsilon^2/r^3$, and then to require $\Omega_{(3)}$ to be closed.  For
this purpose we note that the algebra of the exterior derivative $d$ acting
on the basis $e^i$ is
\bea
de^1&=&-\left(\frac{y}{H}-K\right)e^1\wedge e^2 + 2K e^3\wedge e^4,\nonumber\\
de^2&=&0, \nonumber\\
de^3&=&-e^5\wedge e^4 + Ke^2\wedge e^3 - \frac{y}{H} e^1 \wedge e^4,
\nonumber\\
de^4&=& e^5 \wedge e^3 + K e^2 \wedge e^4 + \frac{y}{H} e^1 \wedge e^3, \nonumber\\
de^5 &=& -e^1\wedge e^2 +e^3\wedge e^4,
\eea
This follows directly from (\ref{de}) and (\ref{eq:ei}).
We find that taking the exterior derivative of $\Omega_3$ produces 14 out
of 15 possible terms.  To simplify matters we assume that we are looking
for a solution that becomes that of \cite{HEK} in the $\epsilon\rightarrow0$
limit, and so we may take that all functions except $\alpha_1$ are of order
$\epsilon^2$ already.  This allows us to drop certain $\epsilon^4$ terms.
The actual computation is straightforward but quite involved, and we present
some of the partial results in appendix \ref{3form}. The most important
result is that such an imaginary self dual 3-form can be constructed
explicitly, resulting in
\begin{eqnarray}
{\Omega}_3&=&\left ( \frac{dr}{r}+i\frac{e_{(s)}^5}{3}\right )
\wedge \biggl[
\frac{3}{2}\frac{1}{(1-cy)^2}
(e_{(s)}^1\wedge e_{(s)}^2+e_{(s)}^3\wedge e_{(s)}^4)\nonumber\\
&&\kern4em+{\alpha}_2(r,y)(e_{(s)}^1\wedge e_{(s)}^4+e_{(s)}^2\wedge e_{(s)}^3)
+i{\alpha}_3(r,y)(e_{(s)}^1\wedge e_{(s)}^3-e_{(s)}^2\wedge
e_{(s)}^4)\biggr]\nonumber\\
&&\kern-2em+b_1(r, y)\left (\frac{e_{(s)}^5}{3}+i\frac{dr}{r}\right)\wedge
\bigl[(e_{(s)}^1\wedge e_{(s)}^3+e_{(s)}^2\wedge e_{(s)}^4)
+i(e_{(s)}^1\wedge e_{(s)}^4-e_{(s)}^2\wedge e_{(s)}^3)\bigr]\nonumber\\
&&\kern-2em
+il_2(r,y)(e_{(s)}^3+ie_{(s)}^4)\wedge
\left(\!\frac{dr}{r}\wedge \frac{e_{(s)}^5}{3}+\frac{1}{6}e_{(s)}^1\wedge e_{(s)}^2\!\right),\nonumber\\
\label{eq:isd3f}
\end{eqnarray}
where
\begin{eqnarray}
{\alpha}_2(r,y)&=&\frac{3}{4}\frac{{\epsilon}^2}{r^3}\left (1-\frac{1}{(1-cy)^4}\right )
+\frac{3}{4}\frac{{\epsilon}^2}{r^3}\frac{1}{c^2{\cal Q}(y)}
\left [(1-ac^2)\left (\frac{1}{(1-cy)^4}-1 \right )
\right. \nonumber \\&&
\left. -\frac{9}{(1-cy)^2}+\frac{11}{1-cy}
+4-6(1-cy)-3(1-cy)^2+3(1-cy)^3
\right],\quad\\
{\alpha}_3(r,y)&=&-\frac{3}{4}\frac{{\epsilon}^2}{r^3}\left (1-\frac{1}{(1-cy)^4}\right )
+\frac{3}{4}\frac{{\epsilon}^2}{r^3}\frac{1}{c^2{\cal Q}(y)}
\left[ (1-ac^2)\left (\frac{1}{(1-cy)^4}-1 \right )
\right. \nonumber \\&&
\left.-\frac{3}{(1-cy)^2} +\frac{5}{1-cy}
-8+6(1-cy)+3(1-cy)^2-3(1-cy)^3
\right],\\
b_1(r,y)&=&\frac{3}{2}\frac{{\epsilon}^2}{r^3}\left [\ln \frac{2r^3}{{\epsilon}^2}-4
+\frac{3}{(1-ac^2)(1-cy)}\right. \nonumber\\&&
\left. +\frac{c}{2(1-ac^2)}\sum_{i=1}^{3}
\frac{(a+2acy_i+(1-ac^2)y_i^2)}{y_i(1-cy_i)}\ln(|y-y_i|)
\right],\nonumber\\
l_2(r,y)&=&\frac{9}{2}\frac{{\epsilon}^2}{r^3}\frac{1}{cH(1-cy)}
\left[(1-cy)-\frac{1}{(1-cy)}\right]^2.
\label{eq:b1ry}
\end{eqnarray}
Here ${\cal Q}(y)=a-3y^2 +2cy^3$ and $y_i$ are the three roots
resulting from ${\cal Q}(y)=0$.

In the absence of a deformation ($\delta=\epsilon=0$) the above reduces
to the simple result for $\Omega_{(3)}$ given by \cite{HEK}
\be
\alpha_1(y)= \frac{3}{2(1-cy)^2}.
\ee
Moreover, in the limit of \cite{HEK} $\Omega_{(3)}$ is a $(2,1)$ form, a fact 
that can be established using the complex coordinates of \cite{MS} as
reviewed in section \ref{kah}. Another interesting case, $c=0$ (when $K=0$)
is the KS \cite{KS} solution with
\be
\alpha_1\to\fft32, \quad \beta_2=\beta_3\to\fft32 \frac{\epsilon^2}{r^3}
\left(\ln\left(\frac{2r^3}{\epsilon^2}\right)-1\right),
\ee
and all other functions set to zero.  For the KS solution the corresponding
$\Omega_3$ is also (2,1).
In both of the above limiting cases, the fact that they are $(2,1)$
guarantees that the solution is supersymmetric. We have not addressed
the question of the precise Dolbeault decomposition of our $\Omega_3$;
however we hope to return to this question in the future.

\subsection{Other forms}

By comparing the imaginary self dual 3-form obtained above with the
Klebanov-Strassler solution \cite{KS} and the Herzog, Ejaz and Klebanov
solution \cite{HEK}, we may extract the NS-NS 3-form flux $H_3$ and the
R-R form $F_3$ from the real and imaginary parts of $\Omega_3$.  We
find
\begin{eqnarray}
\frac{F_3}{M}&=& {\alpha}_1(y)\biggl[ e^1_{(s)}\wedge e^2_{(s)}
\wedge\frac{e^5_{(s)}}{3}+ e^3_{(s)}\wedge e^4_{(s)}\wedge\frac{e^5_{(s)}}{3}\biggr]\nonumber\\&&
+[{\alpha}_2(r,y)+b_1(r,y)] e^1_{(s)}\wedge e^4_{(s)}\wedge\frac{e^5_{(s)}}{3}
+[{\alpha}_2(r,y)-b_1(r,y)] e^2_{(s)}\wedge e^3_{(s)}\wedge\frac{e^5_{(s)}}{3}
\nonumber\\&&
+[{\alpha}_3(r,y)+b_1(r,y)] e^1_{(s)}\wedge e^3_{(s)}\wedge\frac{dr}{r}
+[-{\alpha}_3(r,y)+b_1(r,y)] e^2_{(s)}\wedge e^4_{(s)}\wedge\frac{dr}{r}
\nonumber\\&&
-l_2(r,y)e^3_{(s)}\wedge \frac{e^5_{(s)}}{3}\wedge \frac{dr}{r}
+\frac{1}{6}l_2(r,y) e^1_{(s)}\wedge e^2_{(s)}\wedge e^3_{(s)},\\
\frac{H_3}{g_sM}&=& {\alpha}_1(y)\left[ e^1_{(s)}\wedge e^2_{(s)}\wedge\frac{dr}{r}+ e^3_{(s)}\wedge e^4_{(s)}\wedge\frac{dr}{r}
\right]\nonumber\\&&
+[{\alpha}_2(r,y)-b_1(r,y)] e^1_{(s)}\wedge e^4_{(s)}\wedge\frac{dr}{r}
+[{\alpha}_2(r,y)+b_1(r,y)] e^2_{(s)}\wedge e^3_{(s)}\wedge\frac{dr}{r}
\nonumber\\&&
+[-{\alpha}_3(r,y)+b_1(r,y)] e^1_{(s)}\wedge e^3_{(s)}\wedge\frac{e^5_{(s)}}{3}
+[{\alpha}_3(r,y)+b_1(r,y)] e^2_{(s)}\wedge e^4_{(s)}\wedge\frac{e^5_{(s)}}{3}
\nonumber\\&&
+l_2(r,y)e^4_{(s)}\wedge \frac{e^5_{(s)}}{3}\wedge \frac{dr}{r}
-\frac{1}{6}l_2(r,y) e^1_{(s)}\wedge e^2_{(s)}\wedge e^4_{(s)}.
\end{eqnarray}
The NS-NS 3-form flux can be derived from a potential $H_3=dB_2$ where
\begin{eqnarray}
\frac{B_2}{g_sM}&=& f(r,y)[ e^1_{(s)}\wedge e^2_{(s)}+
e^3_{(s)}\wedge e^4_{(s)}]\nonumber\\
&&+\left[-\ft{1}{3}{\alpha}_3(r,y)+\ft{1}{3}b_1(r,y)-{\delta}f(r,y)
\right ]e^1_{(s)}\wedge e^4_{(s)}\nonumber\\
&&+\left[-\ft{1}{3}{\alpha}_3(r,y)-\ft{1}{3}b_1(r,y)+{\delta}f(r,y)
\right]e^2_{(s)}\wedge e^3_{(s)}\nonumber\\
&&+\ft{1}{3}l_2(r,y)e^3_{(s)}\wedge \frac{dr}{r},
\end{eqnarray}
with
\begin{equation}
f(r,y)=\frac{1}{2}\frac{1}{(1-cy)^2}\left [\ln \left(
\frac{2r^3}{{\epsilon}^2}\right)-1\right].
\end{equation}
If we take the limit (\ref{limit}), then only the term proportional to
$f(r,y)$ survives. This is in agreement with the Klebanov-Strassler expression
for $B_2$ expanded to first order in $\epsilon^2$, which is
\begin{equation}
\frac{B_2}{g_sM}=\frac{1}{2}\left[\ln\left(\frac{2r^3}{{\epsilon}^2}\right)-1
\right]
[e^1_{(s)}\wedge e^2_{(s)}+ e^3_{(s)}\wedge e^4_{(s)}]+{\cal O}({\epsilon}^4).
\end{equation}
Due to self duality of the complex 3-form, the dilaton field is constant,
$\phi=0$. Since $F_{3\;\mu\nu\rho}{H_3}^{\mu\nu\rho}=0$, the axion vanishes
as well.  The five-form flux is
\begin{eqnarray}
F_5&=&{\cal F}_5+*{\cal F}_5,\nonumber\\
{\cal F}_5 &=& B_2\wedge F_3,\nonumber\\
\frac{1}{g_sM^2}{\cal F}_5&=&
\frac{2}{3}f(r,y){\alpha}_1(y)
e^1_{(s)}\wedge e^2_{(s)}\wedge e^3_{(s)}\wedge e^4_{(s)}\wedge e^5_{(s)}
\nonumber\\&&
-\frac{l_2(r,y)}{6(1-cy)^2}\ln \left (\frac{2r^3}{{\epsilon}^2}\right )
e^1_{(s)}\wedge e^2_{(s)}\wedge e^3_{(s)}\wedge e^5_{(s)}\wedge 
\frac{dr}{r}+{\cal O}({\epsilon}^4).
\end{eqnarray}
The effect of the inhomogenous metric is reflected in the five-form at
first order in ${\epsilon}^2$.

\subsection{The warp factor}

The equation for the warp factor can be extracted from the $F_5$ equation
of motion
\be
dF_5=H_3\wedge F_3.
\ee
The form of $F_5$ from (\ref{ansatz}) implies that
\be
d*dh^{-1}\wedge dx^0\wedge dx^1\wedge dx^2\wedge dx^3= H_3\wedge F_3,
\ee
where
\bea
*\calf_5&=&*dh^{-1} \wedge dx^0\wedge dx^1\wedge dx^2\wedge dx^3\nonumber\\
&=&-r^5 h'\,e_{(s)}^1\wedge e_{(s)}^2\wedge e_{(s)}^3 \wedge e_{(s)}^4
\wedge e_{(s)}^5\nonumber\\
&&+r^3 \dot{h} H(y) \, e^r \wedge e_{(s)}^1\wedge e_{(s)}^3 \wedge e_{(s)}^4
\wedge e_{(s)}^5
-r^3 \dot{h} H(y)\,\delta\, e^r \wedge e_{(s)}^1\wedge e_{(s)}^2
\wedge e_{(s)}^3 \wedge e_{(s)}^5,\nonumber\\
\eea
where in the last equation we used the fact that $dy=-H(y)(e_{(s)}^2-\delta\,
e_{(s)}^4)$.

\subsubsection{D3 branes on the deformation}\label{pured3}

We consider first the simple case where there are no fractional D3 branes,
that is, the case of zero three-form flux.  The above equations simplify
schematically to the familiar
\be
\nabla^2 h=\frac{1}{\sqrt{g}}\partial_{m}\left(\sqrt{g}g^{m
n}\partial_{n} h\right)=\delta(\vec{x}\,).
\ee
The first order perturbation is traceless, implying that the first order
correction to $\det g$ is zero.  Further, if we take that the D3 branes
are at the tip of the cone, the zeroth order harmonic function is simply
$A+{B}/{r^4}$.  The metric perturbation has no $r$ indices, and so the
resulting expansion in $\epsilon^2$ of the above equation trivially yields
\be
\nabla^2 h_1=0,
\ee
where $h_1$ is the $\epsilon^2$ correction of $h$.  The first order
correction to the harmonic function can therefore be set to $0$
consistently.

Given this fact, taking the usual limit $N\rightarrow \infty$, with $g_sN$
fixed gives the following geometry
\be
ds^2=R^2\left[ds^2_{AdS_5} + ds^2_{Y^{p,q}} +
\frac{\epsilon^2}{r^3}\left(\frac{\left(e^1e^3+e^2e^4\right)}{3(1-cy)^2}\right)\right].
\ee

As we mentioned before, this deformation of the solution corresponds
to chiral symmetry breaking. To see this, note that the vector
$\partial/\partial \psi$ is no longer a Killing vector for the
perturbed metric.  The new metric therefore also breaks the $U(1)_R$
symmetry (of the gauge theory) associated with (\ref{u1inse})
\cite{sequiver}
\be
3\frac{\partial}{\partial \psi} - \frac{1}{2}\frac{\partial}{\partial \alpha},
\ee
and so chiral symmetry has been broken. Let us see this argument using
standard AdS/CFT arguments. To a supergravity deformation of the form
\be
ar^{\Delta-4}+b r^{-\Delta},
\ee
corresponds a field theory deformation by an operator ${\cal O}$ such
that
\be
{\cal H}\to {\cal H}+a {\cal O}, \qquad \langle{\cal O}\rangle=b.
\ee
Applying this to our situation, we have a deformation that goes as
\be
\epsilon^2 \; r^{-3},
\ee
which implies an expectation value for a dimension-3 operator which should be
of the form
\be
\langle\bar{\Psi}\Psi\rangle=\epsilon^2.
\ee
Thus the gauge theory living in the worldvolume of a stack of D3
branes at the tip of the complex deformation of the cone over
$Y^{p,q}$ corresponds to placing the superconformal quiver gauge
theory in a different vacuum where the operator $\bar{\Psi}\Psi$
has a nonzero vacuum expectation value.

\subsubsection{D3 branes and fractional D3 branes}

The problem of solving for the warp factor in the presence of
fractional D3 branes on $B^{p,q}$ was addressed in \cite{HEK}.
We will take their solution as the zeroth order in epsilon term
(as our solution collapses to theirs in the
$\epsilon \rightarrow 0$ limit).  Curiously, their solution
remains unperturbed, as will be shown here.  First, consider the
resulting equation for $h$,
\be
-\nabla^2_{(6)} h = \frac{1}{6}\left|H_3\right|^2.
\ee
Next, note that in the first order deformed solution none of the indices
(in vielbein basis) on the $\mathcal O(\epsilon^2)$ terms agree with those
of the $\mathcal O(\epsilon^0)$ term.  The basis is diagonal, and so the
zeroth order term with indices contracted with the first order term vanishes.
Therefore, there are no $\mathcal O(\epsilon^2)$ source terms on the right
hand side of the above equation.  As before, the zeroth order solution takes
care of the right hand side, leaving us with only the first order equation
\be
\frac{1}{\sqrt{g}}\partial_{m}\left[\sqrt{g}g^{m n}
\partial_{n} h_1(r,y)\right]=
\frac{1}{\sqrt{g}}\partial_{m}\left[\sqrt{g}h^{m n}
\partial_{n} h_0(r,y)\right],
\ee
where we read the right hand side of the equation as being a
source term.  Without the fractional D3 branes, the source had
vanished because $h_0$ only depended on $r$, and $h^{m n}$
has no $r$ indices.  However, now $h_0$ depends both on $r$
and $y$.  Interestingly, however, the right hand side still vanishes
because
\be
\sqrt{g}h^{m y}\partial_y h_0(r,y)=
\fft{\epsilon^2\sqrt{3}}{54\mathcal Q(y)^{1/2}} \left[\begin{matrix}
0\\
-\cos(\psi)\dot{h}_0(r,y)\sin(\theta)\\
-\sin(\psi)\dot{h}_0(r,y)\\
0\\
-c \cos(\theta) \sin(\psi) \dot{h}_0(r,y)\\
\cos(\theta)\sin(\psi) \dot{h}_0(r,y)
\end{matrix} \right],
\ee
where this has been written in $[r,\theta,\phi,y,\beta,\psi]$ order.  The
divergence of this quantity obviously vanishes, with the first non-zero and
last non-zero terms canceling after the appropriate derivatives are taken.
Thus, again, we may consistently set the first order perturbation to the warp
factor to be zero because the source term is absent.

\section*{Acknowledgments}

We are especially grateful to S.~Franco for various comments and
discussions on topics related to this work. We have enjoyed comments by
C.~N\'u\~nez, J.~Sonnenschein and D.~Vaman. This work is supported by
the US Department of Energy under grant DE-FG02-95ER40899.

\appendix

\section{Gravitational perturbation theory}

In section \ref{deformation}, we have relied heavily on linearized
gravity.  We simply note here the equations to second order.  First,
we write the metric perturbed to second order
\be
\tilde{g}_{m n} = \,^0\!g_{mn}+\,^1\!h_{mn}+\,^2\!h_{mn},
\ee
where the pre-superscripts denote the order of the perturbation.
Given the above decomposition, the following expressions are valid to
second order
\bea
\tilde{g}^{ab} &=& \,^0\!g^{ab}-\,^1\!h^{ab}-\,^2\!h^{ab}+\,^1\!h^{a}_c\,^1\!h^{cb},\nonumber\\
\tilde{\Gamma}^{a}_{bc} &=&
\,^0\!\Gamma^{a}_{bc}+\,^1\!\Gamma^{a}_{bc}
+\,^2\!\Gamma^{a}_{bc}+\,^{(1,1)}\!\Gamma^{a}_{bc},\nonumber\\
\tilde{R}^{a}_{bcd} &=& \,^0\!R^{a}_{bcd}+
\,^1\!\Gamma^{a}_{bd;c}-\,^1\!\Gamma^{a}_{bc;d}
+\,^2\!\Gamma^{a}_{bd;c}-\,^2\!\Gamma^{a}_{bc;d}
+\,^{(1,1)}\!\Gamma^{a}_{bd;c}-\,^{(1,1)}\!\Gamma^{a}_{bc;d} \nonumber
\\ && \;\;\;\; +\,^1\!\Gamma^{a}_{ec}\,^1\!\Gamma^{e}_{bd}
-\,^1\!\Gamma^{a}_{ed}\,^1\!\Gamma^{e}_{bc}.
\eea
We have defined the convenient quantities
\bea
\,^1\!\Gamma^{a}_{bc}&=&\frac{1}{2}\left(\,^1\!h^{a}_{b;c}+\,^1\!h^{a}_{c;b}
-\,^1\!h_{bc}^{\;\; ;a}\right), \nonumber \\
\,^2\!\Gamma^{a}_{bc}&=&\frac{1}{2}\left(\,^2\!h^{a}_{b;c}+\,^2\!h^{a}_{c;b}
-\,^2\!h_{bc}^{\;\; ;a}\right),\nonumber \\
\,^{(1,1)}\!\Gamma^{a}_{bc} &=&
-\frac{1}{2}\,^1\!h^{ad}\left(\,^1\!h_{db;c}
+\,^1\!h_{dc;b}-\,^1\!h_{bc;d}\right).
\eea
In all of the above equations $\,^0\!g^{ab}$ is used to raise and
lower indices, and to construct the Christoffel symbols used in the
covariant derivatives.  Constructing Einstein's equations to second
order is now trivial.  For the purposes of this paper, all first order
perturbations are transverse, traceless and are zero modes of the
Lichnerowicz operator.  In addition, the background is Ricci flat.
This greatly simplifies the linearized Einstein's equations, and they
become
\bea
&&\ft{1}{2}\Bigl[\,^2\!h^a_{b;d;a}+\,^2\!h^a_{d;b;a}
-\,^2\!h_{bd;a}^{\;\;\;\; ;a}-\,^2\!(h^a_a)_{;b;d}\Bigr]\nonumber \\
&&\kern6em=
\ft{1}{2}\Bigl[\,^1\!h^{ac}\left(\,^1\!h_{cb;d;a}+\,^1\!h_{cd;b;a}
-\,^1\!h_{bd;c;a}-\,^1\!h_{ac;b;d}\right)\Bigr]\nonumber \\
&&\kern8em
-\ft{1}{4}\,^1\!h^{ac}_{\;\; ;d}\,^1\!h_{ac;b}
-\ft{1}{2}\Bigl[\,^1\!h_{da;c}\,^1\!h_{b}^{\;\; ba; c}
-\,^1\!h_{da;c}\,^1\!h_{b}^{\;\; bc; a}\Bigr].
\eea
We may read this as the second order term being sourced by the first
order term.  In addition, we may perform a simple check of the above
equation.  In the case when $\,^1\!h$ is zero, the leading
contribution to the metric is $\,^2\!h$.  Therefore, what we have done
should collapse to ``leading order" perturbation theory.  This is
indeed the case, as we can see the Lichnerowicz operator acting on
$\,^2\!h$ on the left hand side of the equation.

\section{Explicit derivation of $\Omega_3$}\label{3form}

Since one of the main results supporting the existence of a chiral
symmetry broken phase for the cascading quiver theory relies on the
existence of the imaginary self dual 3-form $G_3$, we present the details
of its derivation below. In particular, we strive to make clear the
assumptions that go into solving the system.

Imposing
\be
d\Omega_3=0,
\ee
where $\Omega_3$ is given in (\ref{eq:omeg3}) results in the following
system of equations for the ten quantities $\alpha_i,\beta_i,\lambda_i$:
\begin{enumerate}
\item $r\alpha _1'+r\beta _1'+H\dot{\lambda}_1-\lambda_1\left(\frac{y}{H}-K\right)=0,$
\item $-r\alpha_3'-3\delta\alpha_1 -3\alpha_2 +r\beta_3'+3\beta_2+\frac{y}{H}\lambda_4=0,$
\item $\delta\left(r\alpha_1'-3\alpha_1\right)+r\alpha_2' +3\alpha_3+r\beta_2'+3\beta_3-\frac{y}{H}\lambda_3=0,$
\item $-\delta\left(r\alpha_1'-3\alpha_1\right)+r\alpha_2' +3\alpha_3-r\beta_2'-3\beta_3-H\dot{\lambda}_3+K\lambda_3=0,$
\item $r\alpha_3'-3\delta\alpha_1 +3\alpha_2 +r\beta_3'+3\beta_2-H\dot{\lambda}_4+K\lambda_4=0,$
\item $r\alpha_1'-r\beta_1'+2K\lambda_1=0,$
\item $-H\dot{\alpha}_3+2\frac{y}{H}\alpha_3 -H\dot{\beta}_3+\frac{1}{3}\lambda_3-\frac{1}{6} r\lambda_4'=0,$
\item $-H\dot{\alpha}_1\delta-4K\delta\alpha _1 -H\dot{\alpha}_2+2\frac{y}{H}\alpha _2+H\dot{\beta}_2-
\frac{1}{6} r\lambda _3'+\frac{1}{3} \lambda _4 =0,$
\item $-H\dot{\alpha_1}+4K\alpha_1-H\dot{\beta}_1+\frac{1}{6} r\lambda_1'+\frac{1}{3}\lambda_2 =0,$
\item
$H\dot{\alpha}_3-2\frac{y}{H}\alpha_3-H\dot{\beta}_3-\frac{1}{2} \lambda_3=0,$
\item
$-H\dot{\alpha}_1\delta-4K\delta\alpha_1 -H\dot{\alpha}_2+2\frac{y}{H}\alpha_2-H\dot{\beta}_2+\frac{1}{2} \lambda_4=0,$
\item $-H\dot{\alpha}_1+4K\alpha_1+H\dot{\beta}_1 =0,$
\item $\frac{2}{3}{\beta}_1 -\frac{1}{6}\left[H\dot{\lambda}_2-{\lambda}_2
\left( \frac{y}{H}+K\right)\right]=0,$
\item $\frac{1}{3} \lambda _1 + \frac{1}{6} r\lambda_2'=0,$
\end{enumerate}
where dot means derivative with respect to $y$ and prime with respect
to $r$. Note that we have already taken into account the expansion to
leading order in $\epsilon^2$.  In particular, all quantities except
$\alpha_1$ are viewed as $\mathcal O(\epsilon^2)$ terms.  This is why
$\delta$ (which is of order $\epsilon^2$) only survives in combination
with $\alpha_1$.

These equations may be simplified by defining the new variables
\begin{eqnarray}
B_1=\ft12({\beta}_2+{\beta}_3),&&
B_2=\ft12({\beta}_2-{\beta}_3),\nonumber\\
l_1=\ft12({\lambda}_3+{\lambda}_4),&&\;\,
l_2=\ft12({\lambda}_3-{\lambda}_4),\nonumber\\
A_1=\ft12({\alpha}_2+{\alpha}_3),&&
A_2=\ft12({\alpha}_2-{\alpha}_3),
\end{eqnarray}
and by taking appropriate linear combinations.

The result is a set of equations involving only ${\beta}_1$, ${\lambda}_1$,
${\lambda}_2$ and ${\alpha}_1$:
\begin{enumerate}
\item $2r{\beta}_1'+H\dot{\lambda}_1-\left (K+\frac{y}{H} \right ) {\lambda}_1=0,$
\item $2r{\alpha}_1'+H\dot{\lambda}_1 +\left ( 3K-\frac{y}{H}\right ) {\lambda}_1=0,$
\item $-12H\dot{\beta}_1+r{\lambda}_1'+2{\lambda}_2=0,$
\item $-H\dot{\alpha}_1+4K\alpha_1+H\dot{\beta}_1=0,$
\item $4{\beta}_1-H\dot{\lambda}_2+\left ( K+\frac{y}{H}\right ){\lambda}_2=0,$
\item $2 \lambda _1 + r\lambda _2'=0,$
\end{enumerate}
a second set involving $B_1$, $l_2$, $A_2$ and ${\alpha}_1$:
\begin{enumerate}\addtocounter{enumi}{6}
\item $2rB_1'+6B_1+H\dot{l}_2-\left (K+\frac{y}{H} \right ) {l}_2=-\delta(r{\alpha}_1'-6{\alpha}_1),$
\item $2rA_2'-6A_2-H\dot{l}_2+\dot{H}l_2=0,$
\item $12H\dot{B}_1+l_2-rl_2'=0,$
\item $4H\dot{A}_2-\frac{8y}{H}A_2+\frac{5}{3}l_2+\frac{1}{3}rl_2'=
-16K\delta{\alpha}_1,$
\end{enumerate}
and finally a third set involving only $B_2$, $l_1$, $A_1$ and ${\alpha}_1$:
\begin{enumerate}\addtocounter{enumi}{10}
\item $2r A_1'+6A_1-H\dot{l}_1+\dot{H}l_1=0,$
\item $-2rB_2'+6B_2-H\dot{l}_1+\left (K+\frac{y}{H} \right ) {l}_1=\delta r {\alpha}_1',$
\item $-12H\dot{B}_2+l_1+rl_1'=0,$
\item $-4H\dot{A}_1+\frac{8y}{H}A_1+\frac{5}{3}l_1-\frac{1}{3}rl_1'=
16K\delta{\alpha}_1.$
\end{enumerate}

Furthermore, since $\alpha_1$ only enters the second and third sets
of equations in conjunction with $\delta$, the $\mathcal O(\epsilon^2)$
part of $\alpha_1$ decouples from those sets of equations (the zeroth
order part acts as a source).  Hence if one is given the zeroth order
form for ${\alpha}_1$, then these sets of equations decouple. A good
guess can be made by looking at the limiting cases of the
Klebanov-Strassler \cite{KS} and Herzog, Ejaz and Klebanov \cite{HEK}
solutions.  Based on this, we take ${\alpha}_1$ to be given by
\begin{equation}
{\alpha}_1=\frac{3}{2}\frac{1}{(1-cy)^2}
\end{equation}
This choice of ${\alpha}_1$, together with ${\lambda}_1={\lambda}_2
={\beta}_1=0$ then satisfies the first set of equations without any
correction at $\mathcal O(\epsilon^2)$. Next, taking the $r$-dependence of
$A_1$ to be $r^{-3}$ and setting $B_2=l_1=0$ is consistent with third set
of equations. Equation 14 is then solved to give
\begin{equation}
A_1=\frac{3}{4}\frac{{\epsilon}^2}{r^3}\frac{1}{c^2{\cal Q}(y)}\left[
\frac{1-ac^2}{(1-cy)^4}-\frac{6}{(1-cy)^2}+\frac{8}{1-cy}-3+ac^2
\right],
\end{equation}
where ${\cal Q}(y)=a-3y^2+2cy^3$. The integration constant is fixed such that one gets the Klebanov-Strassler solution in the limit (\ref{limit}).

The remaining second set of equations present more of a challenge.
To check their consistency, we make an Ansatz for $r$ dependence as follows:
\begin{eqnarray}
B_1&=&\frac{3{\epsilon}^2}{2r^3}\left [{\mu}(y)\ln \frac{r^3}{{\epsilon}^2}
-{\nu}(y)\right],\nonumber\\
l_2&=&\frac{H}{r^3}\left [{\theta}(y)\ln \frac{r^3}{{\epsilon}^2}
-{\psi}(y)\right],\nonumber\\
A_2&=&\frac{1}{H^2r^3}\left [{\rho}(y)\ln \frac{r^3}{{\epsilon}^2}
-{\tau}(y)\right].
\end{eqnarray}
The system now reduces to eight equations for six functions of $y$:
\begin{enumerate}
\item $9{\epsilon}^2\dot{\mu}+2\theta=0,$
\item $18{\epsilon}^2\dot{\nu}+4\psi+3\theta=0,$
\item $H^2\dot{\theta}-2y\theta =0,$
\item $9{\epsilon}^2\mu+2y\psi-H^2\dot{\psi}=9\frac{{\epsilon}^2}{(1-cy)^4},$
\item $12{\rho}+H^4\dot{\theta}=0,$
\item $H^4\dot{\psi}+12\tau+6\rho=0,$
\item $2\dot{\rho}-4\frac{K}{H}\rho +\frac{1}{3}H^2\theta =0,$
\item $-\dot{\tau}+\frac{2K}{H}\tau -\frac{1}{6}H^2\psi +\frac{1}{4}H^2\theta +6KH\frac{{\epsilon}^2}{(1-cy)^4}=0.$
\end{enumerate}
However, we now see that Equation 7 can be obtained from Equations 3 and 5.
and Equation 8 can be obtained from Equation 4 with the help of Equations 6,
1 and 7. This leaves us with six equations for six functions. They can be
solved to give
\begin{eqnarray}
\theta =\rho &=&0,\nonumber\\
\mu &=&1,\nonumber\\
\psi &=&-\frac{27}{2}\frac{{\epsilon}^2}{c{\cal Q}(y)}
\left [(1-cy)-\frac{1}{(1-cy)}\right ]^2,\nonumber\\
\tau &=&\frac{3}{4}H^2{\epsilon}^2
\left [ \frac{1}{(1-cy)^4}-1\right ]
+\frac{3{\epsilon}^2}{4}\frac{y}{c(1-cy)}
\left [(1-cy)-\frac{1}{(1-cy)}\right ]^2,\nonumber\\
\nu &=&-\frac{2}{9{\epsilon}^2}\int\psi dy\nonumber\\
&=&4-\ln 2 -\frac{3}{(1-ac^2)(1-cy)}\nonumber\\
&&-\frac{c}{2(1-ac^2)}\sum_{i=1}^{3}
\frac{(a+2acy_i+(1-ac^2)y_i^2)}{y_i(1-cy_i)}\ln(y-y_i),
\end{eqnarray}
where $y_i$ are the three roots of the cubic equation ${\cal Q}(y)=0$. The integration constants are chosen such that Klebanov-Strassler solution is obtained in the limit (\ref{limit}).  The resulting imaginary self dual
3-form is given in section \ref{warp} in
Eqns.~(\ref{eq:isd3f})--(\ref{eq:b1ry}).


\end{document}